\documentclass[conference, a4paper]{IEEEtran}

\usepackage{cite}

\ifCLASSINFOpdf
  \usepackage[pdftex]{graphicx}
  \graphicspath{{../pdf/}{../jpeg/}}
  \DeclareGraphicsExtensions{.pdf,.jpeg,.png}
\else
\fi
\usepackage{float}
\usepackage[cmex10]{amsmath}
\usepackage{amsthm}
\usepackage{amssymb}
\usepackage[caption=false,font=normalsize,labelfont=sf,textfont=sf]{subfig}
\usepackage{graphicx}

\usepackage{url}

\usepackage{bm}

\newcommand{\R}{\mathbb{R}}
\newcommand{\mb}[1]{\mathbf{#1}}

\usepackage{graphicx}
\usepackage{tabulary}
\usepackage{amsthm}

\newtheorem{thm}{Theorem}
\newtheorem{defn}[thm]{Definition}

\begin{document}
\bstctlcite{IEEEexample:BSTcontrol}

%
\title{Regression Driven F--Transform and Application\\
to Smoothing of Financial Time Series}

\author{\IEEEauthorblockN{Luigi Troiano}
\IEEEauthorblockA{University of Sannio\\ Department of Engineering\\ Benevento, Italy\\
Email: troiano@unisannio.it}
\and
\IEEEauthorblockN{Pravesh Kriplani}
\IEEEauthorblockA{University of Sannio\\ CISELab\\ Benevento, Italy\\
Email: pravesh.kriplani@ciselab.org}
\and
\IEEEauthorblockN{Irene D\'{\i}az}
\IEEEauthorblockA{University of Oviedo\\ Computer Science Department\\ Oviedo, Spain\\
Email: sirene@uniovi.es}
}


%


\maketitle

\begin{abstract}

In this paper we propose to extend the definition of fuzzy transform in order to consider an interpolation of models that are richer than the standard fuzzy transform. We focus on polynomial models, linear in particular, although the approach can be easily applied to other classes of models. As an example of application, we consider the smoothing of time series in finance. A comparison with moving averages is performed using NIFTY 50 stock market index. Experimental results show that a regression driven fuzzy transform (RDFT) provides a smoothing approximation of time series, similar to moving average, but with a smaller delay. This is an important feature for finance and other application, where time plays a key role.

\end{abstract}


%
\IEEEpeerreviewmaketitle

\section{Introduction}

Fuzzy transform (F--transform) \cite{Perfilieva:2006} is a functional tool used to compress and reconstruct information, so to offer a regularized version of the original input. When applied to time series \cite{StepnickaPNPVT09}, it performs a smoothing of data \cite{HOLCAPEK201169, Troiano201211, Troiano2011121, Troiano2010379}. It has been applied to several problems of practical interest (e.g., see references \cite{Gaeta2015, Gaeta2016, Vajgl2012, Loia2017, Tomasiello20161, Zadeh2014}). In \cite{PERFILIEVA20113}. Perfilieva et al. generalized the concept of F-transform to a higher order $F^m (m \geq 0)$, whose components are polynomials of degree $m$. In addition, and prove that the higher $m$, the higher the quality of the approximation.

In Finance, a common practice used to smooth time series is based on moving averages. This approach considers a look-back window used to compute the average. By shifting the window ahead, the moving average changes keeping memory of past values. Although computationally inexpensive and semantically intuitive, a moving average suffers of the lag entailed by the look-back window. For this reason, besides the simple MA, other schemes have been proposed. Among them, the exponential moving average (EMA), that gives more relevance to most recent values.

F-transform is able to offer a well smoothed but better-fitted series of data points. Standard F-transform computes a sequence of values, each offering a simple (constant) model, whose validity is limited to the fuzzy set to which it belongs. Richer models may offer the opportunity to better describe the data. In this paper, Similarly to what has been proposed in \cite{HOLCAPEK2014}, we propose a generalization of standard F--transform in order to include any class of regression models and to offer an example of application to financial time series. The paper is structured as follows. Section \ref{sec:pre} briefly describes some related work. The model is described in Section \ref{sec:method}. Section \ref{sec:exam} shows an example of application to time series in Finance, and finally, in Section \ref{sec:conc} some conclusions are drawn.
     
\IEEEpubidadjcol

\section{Preliminaries and related work}\label{sec:pre}

There are many different smoothing techniques in financial tools. Among them, the moving average is a trend following device. 

Moving average is a calculation to analyze data points by creating series of averages of different subsets of the full data set. There are different moving averages. Among them, Simple, Cumulative, Weighted or Exponential \cite{Kresta2015364}.

Its purpose is to signal that a new trend has either begun or ended or reversed.  It tracks the progress of a trend as a smoothing device. The data is averaged and a smoother line is produced. Therefore the underlying trend becomes easier to view as moving averages line lags the market actions. Shorter averages can reduce the time lags, but the time lags can never be eliminated. In exponentially moving average greater weights are assigned to the recent data. In addition, closing price is generally used for moving average calculations. There are a lot of works based on moving average for studying financial trends (see, for example, \cite{Barrow20166088, Sobreiro201686, Liu20171778}).

Other approach frequently applied to study trends is the fuzzy transform. In the following Fuzzy transform as well as Generalized Fuzzy transform are defined as the proposal detailed in this work is a new generalization of Fuzzy transform, called Regression Fuzzy Transform.

\begin{defn}
Let $[a,b] \subset \mathbb{R}$ be an interval and let $a=x_1,x_2,\ldots ,x_n =b\in \R$ be a set of points, called \emph{nodes}, with $x_i < x_{i+1}$ and $n\geq2$. A fuzzy Ruspini partition \cite{compint} over the interval $[a,b]$ is a collection of fuzzy sets $A_1,\ldots,A_n$ such that for any $i=1 \dots n$,
\begin{itemize}
\item $A_i:[a,b] \rightarrow [0, 1]$, $A_i(x_i) = 1$
\item $A_i(x)=0$ if  $x \in [a, x_{i-1}]\cup [x_{i+1},b]$
\item $A_i(x)$ is continuous
\item $A_i(x) > 0$ if  $x \in (x_{i-1}, x_{i+1})$
\item $A_i(x) \leq A_i(x')$ for any $x,x' \in [x_{i-1}, x_{i}], x < x'$ 
\item $A_i(x) \geq A_i(x')$ for any $x,x' \in [x_{i}, x_{i+1}], x < x'$
\item $\sum\limits_{i=1}^n A_i (x) = 1, \forall x \in [a,b]$
\end{itemize}
\end{defn}
The fuzzy sets  ${A_1, A_2,\ldots , A_n }$ are called \emph{basic functions} \cite{PERFILIEVA200836}.

The fuzzy partition can be made of common hat-shaped (triangular) basic functions, given by
\begin{equation}
A_j(x)=\left\{\begin{array}{rr}
{(x_{j+1}-x)/ (x_{j+1}-x_j),}& {x \epsilon [x_j, x_{j+1}]}\\
{(x - x_{j-1})/(x_j - x_{j-1}),}& {x \epsilon [x_{j-1}, x_j]}\\
{0,}& {otherwise}
\end{array}\right.
\end{equation}
or z-shaped basic functions, such as
\begin{equation}
A_j(x)=\left\{\begin{array}{rr}
{{1\over 2} \left(\cos(\pi{{x-x_j}\over{x_{j+1}-x_j}})+1\right),}& {x \epsilon [x_j, x_{j+1}]}\\
{{1\over 2} \left(\cos(\pi{{x-x_j}\over{x_j-x_{j-1}}})+1\right),} & {x \epsilon [x_{j-1}, x_j]}\\
{0,} & {otherwise}
\end{array}\right.
\end{equation}

According to \cite{Perfilieva:2006} the Fuzzy transform is defined as follows. 
\begin{defn}
Let $f:\R \to \R$ be a continuous function defined in $I$. The fuzzy transform (F--transform) of a function $f(x)$ with respect to the partition $\{A_1, A_2,\ldots , A_n \}$ is the $n$--tuple $[F_1,F_2,\ldots,F_n]$ whose components are
\begin{equation}
F_i={{\int_a^b f(x) A_i(x)dx}\over{\int_a^b A_i(x)dx}}.
\label{eq1}
\end{equation}
\end{defn}
The fuzzy transform offers a minimal solution to the error functional
\begin{equation}
\Phi=\int_a^b(f(x)-F_i)^2 A_i(x)dx.
\label{eq_phi}
\end{equation}

The original function $f$ can be approximately reconstructed
from its fuzzy transform \cite{PERFILIEVA200836} through the inverse F-transform of $f$ with respect to $\{A_1, A_2,\ldots , A_n \}$, that is defined as
\begin{equation}
f_{F,n}=\sum\limits_i^n F_i A_i(x),\qquad x\epsilon I
\label{eq2}
\end{equation}
It offers an approximation of $f$ with arbitrary precision, as stated by Theorem 2 in \cite{Perfilieva:2006}.

In finance and other applications of practical interest, $f$ is given as time series, so that the function $f$ is known only at points $\{p_1,p_2,\ldots,p_s\}$, with $s \gg n$. In this case we refer to the discrete F--transform, and Eq.\eqref{eq1} is replaced by 
\begin{equation}
F_i={{\sum\limits_{j=1}^s f(p_j) A_i(p_j)}\over{\sum\limits_{j=1}^s A_i(p_j)}}, \qquad i=1,\ldots,n
\label{eqfd}
\end{equation}

Accordingly, the inverse F--transform is defined as
\begin{equation}
f_{F,n}(p_j)=\sum_i^n F_i A_i(p_j) \qquad j=1,\ldots,s
\label{eqfi}
\end{equation}

$F^m-$transform is a generalization of $F-$transform where the components are polynomial of degree $m$ instead of constants (that are in fact polynomials with degree $m=0$) (see reference \cite{Perfilieva:2006}).

\begin{defn}\cite{PERFILIEVA20113}
Let $f:[a,b] \to \R$ be a continuous function from $L_2(A_1, \dots A_n)$, and $m\geq 0$ a fixed integer. The $n-$tuple $(F_1^m,\dots,F_n^m)$ is an $F^m-$transform of $f$ with regard to the fuzzy partition $\{A_1, A_2,\ldots , A_n \}$, with $F_k^m$ the $k-th$ orthogonal projection of $f|_{[x_{k-1},x_{k+1}]}$ on $L_2^m(A_k), k=1,\dots,n$.
$L_2^m(A_k)$ is the set of square-integrable functions $f:[x_{k-1},x_{k+1}]\rightarrow \R$ and $L_2(A_1, \dots A_n)$ the set of functions $f:[a,b]\rightarrow \R$ such that for all $k=1,\dots,n$, $f|_{[x_{k-1},x_{k+1}]} \in L_2^m(A_k)$
\end{defn}

The $F^m$ transform is noted $F^m[f]=(F_1^m,\dots,F_n^m)$ with $F_k^m=c_{k,0}P_k^0+c_{k,1}P_k^1+\dots+c_{k,m}P_k^m$.  

${P_k^1,\dots,P_k^m}$ is an orthogonal polynomial system in $L_2(A_k)$. In addition $c_{k,j}, j={1,\dots,m}$ are the coefficients obtained using the inner product $\langle . \rangle_k$ as 

\begin{equation}
c_{k,j}=\frac{\langle f, P_k^j \rangle_k}{\langle P_k^j, P_k^j \rangle_k}= \frac{ \int_a^b f(x) {P_k^j(x) A_k(x)dx}} {\int_a^b {P_k^j(x)P_k^j(x)A_k(x)dx}}.
\end{equation}

For more details about $F^m-$transform refer to \cite{PERFILIEVA20113}. 
Note that the $F-$transform component $F_i$ can be interpreted as a regression model of $f$ whose validity is shading by moving away from the node $x_i$. The inverse F--transform performs an interpolation of models by means of weighted mean according to the validity of each model. In particular, as it was highlighted before, models provided by the standard F--transform can be regarded as polynomial models of order 0. This offers the possibility of generalize the F--transform in order to include any class of regression models. Next section develops such generalization.

\section{Regression driven $F-$transform}\label{sec:method}
F--transform of function $f$ is defined in this case as a regression with respect to the partition $A_1,\ldots,A_n$, the collection of models that arise by performing a regression analysis over each set $A_i$. The degree of membership is used to weigh data points. Therefore points out of the $A_i$'s support do not take part to the regression analysis associated to it.

Regression analysis is performed by assuming a model structure, e.g., a polynomial of a given order, whose parameters has to be estimated in order to fit data that falls within the support of $A_i$.  



Therefore, the result of the analysis for the component $A_i$ is a regression model $R$, such that
\begin{equation}\label{regression}
y_{j:i} = R(p_{j:i},\bm{\beta}_i) + \delta_{j:i} \qquad j = 1..m
\end{equation}
where $p_{j:i} \in supp(A_i)$, $y_{j:i}$ is the corresponding value of function $f$ at point $p_{j:i}$, $R(p_{j:i},\bm \beta_i)$ is the model response at $p_{j:i}$, given the model parameters $\bm \beta_i$, and $\delta_{j:i}$ is the residual.

Model parameters $\bm \beta_i$ can be obtained by means of the \emph{least square} method, minimizing the sum of the squared residuals
\begin{equation}
S_i=\sum_{j=1}^m \delta_{j:i}^2 = \sum_{i=1}^n (y_{j:i} - R(p_{j:i},\bm \beta_i))^2
\end{equation}
\begin{equation}
\hat{\bm \beta}_i=\underset{\bm \beta_i}{\mathrm{argmin}} \;S_i
\end{equation}

In the case of polynomial models, we have
\begin{equation}
y_{j:i} = \beta_{i,0} + \beta_{i,1} p_{j:i} + \beta_{i,2} p_{j:i}^2 + \ldots + \beta_{i,r} p_{j:i}^r + \delta_{j:i}  
\end{equation}
that can be expressed as
\begin{equation}
\mb y_i = \mb X_i \bm \beta_i^\top
\end{equation}
where $\mb y_i$ is the vector representing the function values $y_{j:i}$ of interest for $A_i$ and $\mb X_i$ is a matrix by repeating the vectors $[1, p_{j:i}, p_{j:i}^2, \ldots, p_{j:i}^r]$ at each row, and $\bm \beta_i = [\beta_{i,0}, \beta_{i,1}, \beta_{i,2}, \ldots, \beta_{i,r}]$. In this case, the solution is unique and given as
\begin{equation}
\hat{\bm \beta}_i = (\mathbf{X}_i^\top \mathbf{X}_i)^{-1} \mathbf{X}_i^\top \mathbf{y}_i 
\end{equation}
where $(\mathbf{X}_i^\top \mathbf{X}_i)^{-1} \mathbf{X}_i$ is the Moore-–Penrose pseudoinverse matrix.


The method of ordinary least squares assumes that each data point provides equally precise information. When data points are assumed by a different degree, as in the case of fuzzy sets $A_i$, we can adopt the \textit{weighted least squares} method \cite{Carroll88}. In this case we attempt to minimize
\begin{equation}\label{eq:weighted}
S_i =\sum\limits_{j=1}^n (y_{j:i} - R(p_{j:i}, \bm \beta_i))^2 A_i(p_{j:i})
\end{equation}
In this case, the solution is given as
\begin{equation}
\hat{\bm \beta}_i = (\mathbf{X}_i^\top \mb A_i \mathbf{X}_i )^{-1} \mathbf{X}_i^\top \mb A_i \mathbf{y}_i
\end{equation}
where $\mb A_i$ is the diagonal matrix given by values $A_i(p_{j:i})$.


Until now we considered regression models that are linear in the parameters $\bm \beta_i$. We can extend the application of least-squares to the case of models that are not linear in the parameters. In this case, there is no analytical solution to the problem, so non-linear least-squares adopt an iterative procedure. The basis of this method is to approximate the model by a linear model and then to refine the parameters iteratively. Let $\mb J_i$ the Jacobian matrix, whose elements are defined  as

\begin{equation}
J_{jh:i}=\frac{\partial R(p_{j:i},\bm \beta_i)}{\partial \beta_{h:i}}
\end{equation}
where $\beta_{h:i}$ is the $h$-th element of $\bm \beta_i$. Then, we obtain 

\begin{equation}
\Delta \hat{\bm \beta}_i = (\mathbf{J}_i^\top \mb A_i \mathbf{J}_i )^{-1} \mathbf{J}_i^\top \mb A_i \Delta \mathbf{y}_i
\end{equation}
that is the variation to apply to model parameters  $\bm \beta_i$ at each step.

Algorithms for non-linear least squares estimation include Newton's method, the Gauss-Newton algorithm and the Levenberg-Marquardt method \cite{Bjorck96}.

When the regression model is polynomial and $r = 0$, the unique model parameter is $\beta_i = F_i$, so we get the ordinary F--transform as specific case of a regression driven F--transform (RDFT). As a step further, we can assume the regression model to be linear in the data points, so that as result of regression we get the two model parameters $\bm \beta_i = [\beta_{0:1} \beta_{1:1}]^\top$. More in general, RDFT is defined by means of model parameters $\bm \beta = [\bm \beta_1, \ldots, \bm \beta_n]$. The inverse transform is given by

\begin{equation}
f_{\bm \beta, n}(p_j) = \sum_{i=1}^n R(p_j,\bm \beta_i) A_i(p_j)
\end{equation}




\section{Application to Smoothing of Time Series}\label{sec:exam}
In this section we apply RDFT to smoothing financial time series.
The dataset used is the closing price of NIFTY 50 index in the period from 1 January 2008 to 29 December 2009.\footnote{Available at \url{https://www.nseindia.com/live_market/dynaContent/live_watch/equities_stock_watch.htm}} The period include the subprime mortgage crisis in 2008, characterized by high volatility.

We compare the smoothed series obtained by means of different inverse fuzzy transforms to that obtained by a standard smoothing technique based on moving average (MA). In particular we considered the following cases:
\begin{itemize}
\item Standard IFT over a hat-shaped partition (IFT9)
\item Inverse RDFT, with linear regression, over a hat-shaped partition (IRDFT9-LH)
\item Inverse RDFT, with cubic regression, over a hat-shaped partition (IRDFT9-CH)
\item Inverse RDFT, with linear regression, over a z-shaped partition (IRDFT9-LZ)
\item Inverse RDFT, with cubic regression, over a z-shaped partition (IRDFT9-CZ)
\end{itemize}
The moving average is computed over a period of 30 days (MA30), the fuzzy transform over a partition made of 9 equidistant nodes. That makes the smoothed series comparable.

\begin{figure}[h!]
\includegraphics[trim = 3.25cm 1cm 1cm 0.1cm, clip, width = 9cm]{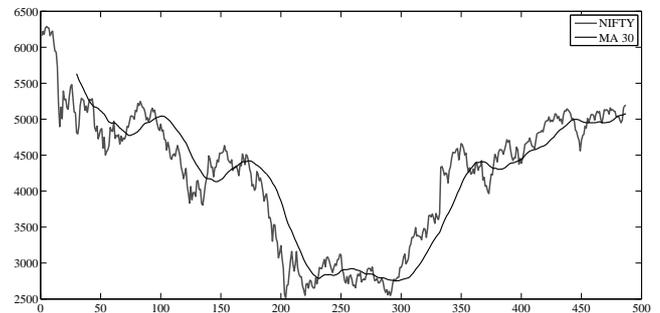}
\caption{Plot of MA30 superimposed to NIFTY 50}
\label{ma}
\end{figure}

\begin{figure}[h!]
\includegraphics[trim = 3.25cm 0.1cm 1cm 0.1cm, clip, width = 9cm]{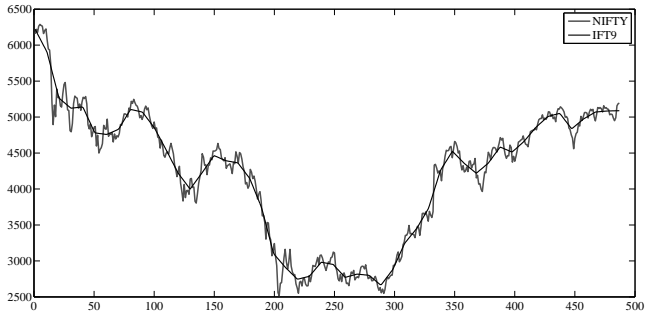}
\caption{Plot of IFT9 superimposed to NIFTY 50}
\label{ift9}
\end{figure}

Smoothing obtained by means of the moving average is shown in Figure~\ref{ma}, where MA30 is superimposed to the NIFTY 50 time series. The smoothing is characterized by a lag due to look-back window used in the moving average computation.
Figure~\ref{ift9} outlines the plot of IFT9 superimposed to NIFTY 50. In this case, the smoothing provided by IFT9 is much closer the original series, still providing a considerable regularization of data.

\begin{figure}[t!]
\includegraphics[trim = 3.25cm 0.1cm 1cm 0.1cm, clip, width = 9cm]{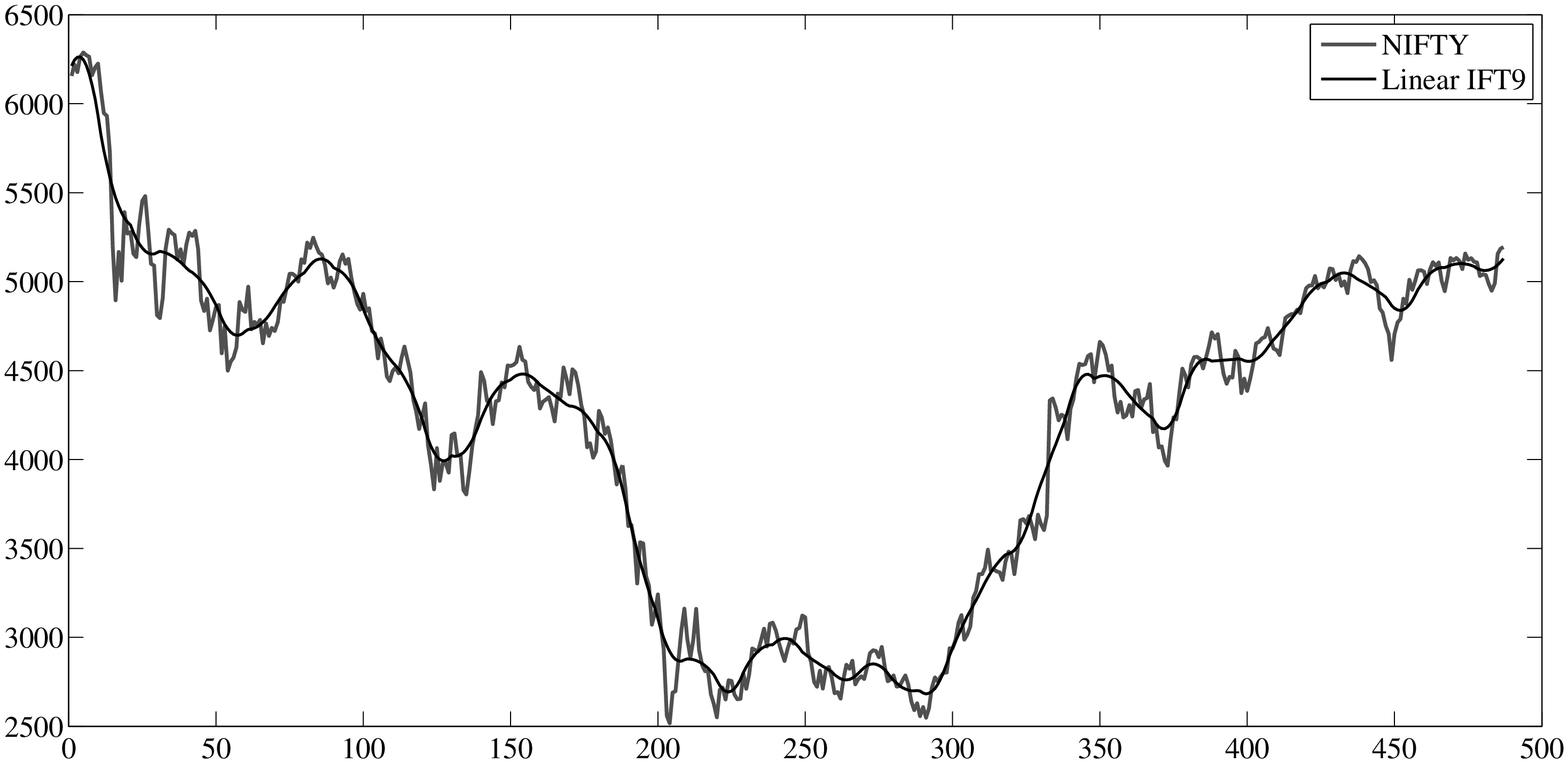}
\caption{Plot of IRDFT9-LH superimposed to NIFTY 50}
\label{linear_ift9}
\end{figure}

As discussed in Section III, the standard F--transform and its inverse can be regarded as 0-degree polynomial regression driven, so that the interpolation performed during reconstruction performs an interpolation of constants $F_i$. By using higher degree polynomials, such as linear ($r=1$) or cubic ($r=3$), we obtain a sequence of local models that better fit data. Figure~\ref{linear_ift9} plots the smoothing obtained by means of IRDFT9-LH (linear regression) superimposed to NIFTY 50.

\begin{figure}[t!]
\includegraphics[trim = 3.25cm 0.1cm 1cm 0.1cm, clip, width = 9cm]{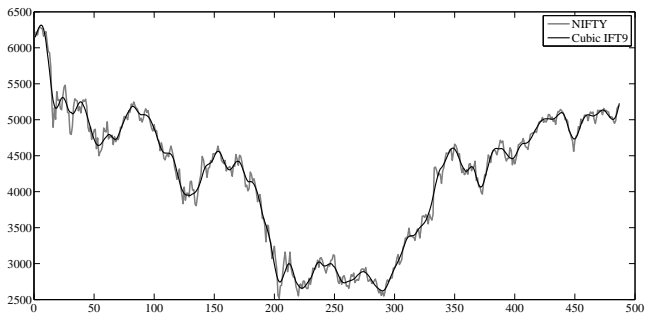}\label{linear_cubic_ift9}
\caption{Plot of IRDFT9-CH superimposed to NIFTY 50}
\end{figure}

The fitness can be further improved, keeping a good level of smoothing, by means of cubic regression, as outlined by Figure~\ref{linear_cubic_ift9}, where IRDFT9-CH is superimposed to NIFTY 50.

\begin{figure}[h!]
\includegraphics[trim = 3.5cm 1.2cm 1cm 0.1cm, clip, width = 9cm]{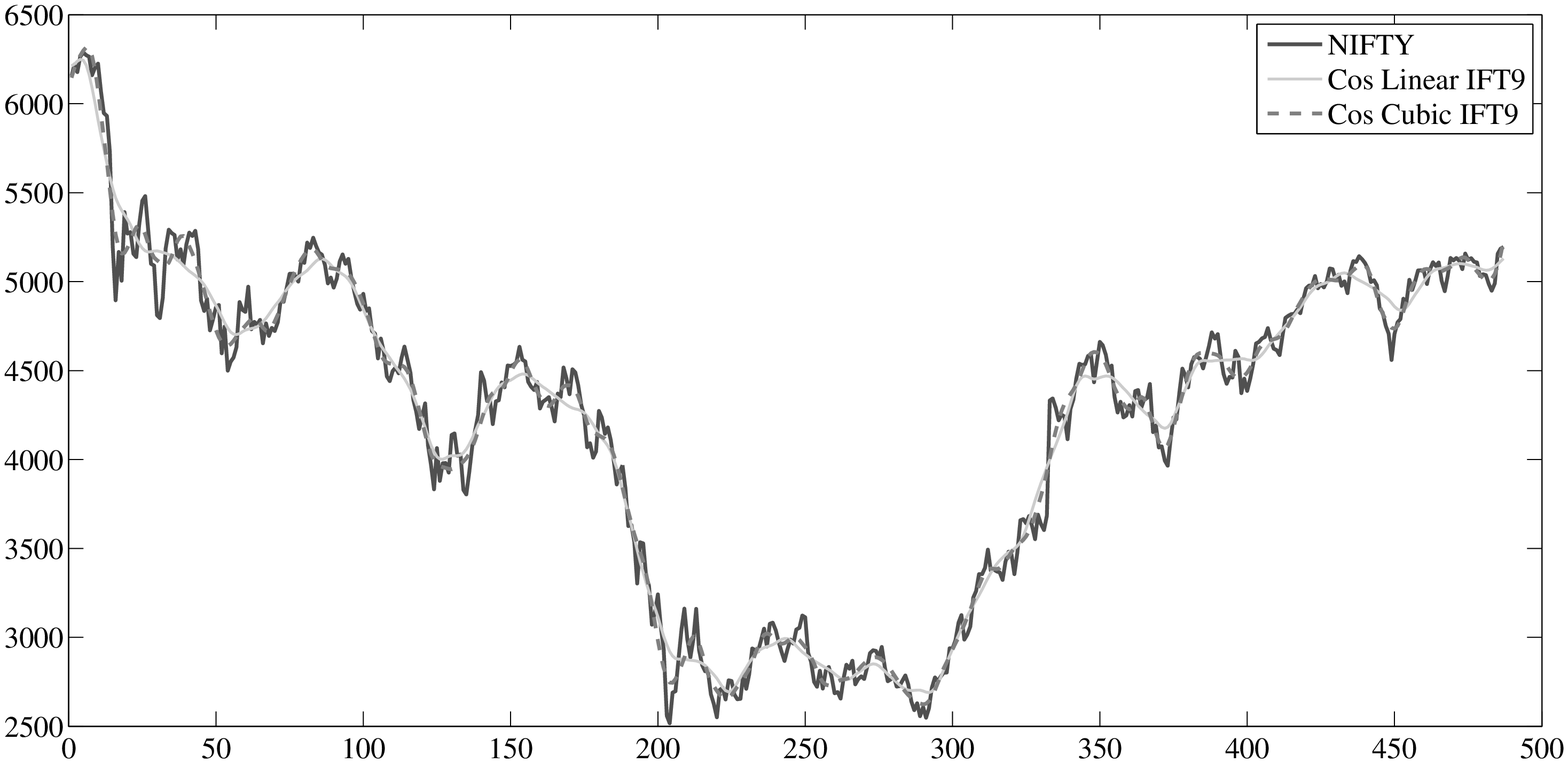} \label{linear_cub_cos_ift9}
\caption{Plot of IRDFT9-LZ and IRDFT9-CZ superimposed to NIFTY50}
\end{figure}

The use of z-shaped functions gives more relevance to the points in the neighborhood of nodes. Figure~\ref{linear_cub_cos_ift9} plots the signals IRDFT9-LZ (linear regression) and IRDFT9-CZ (cubic regression) compared to NIFTY 50.

\begin{table}[t!]
\begin{center}
\label{tab:DR_Val}
\caption{Standard Deviation, Mean Absolute Error and Mean Square Error on the basis of Daily Returns}
\renewcommand{\arraystretch}{2}
\begin{tabular} {|c|c|c| } 
	\hline
	 & Standard Deviation & Mean Absolute Error  \\
	\hline
    MA30 & 19.6931 & 2.578 \\
	\hline
    IRDFT9-LH & 26.4790 & 0.2487 \\
	\hline
    IRDFT9-CH & 37.0529 & 0.0887 \\
    \hline
    IRDFT9-LZ & 26.0214 & 0.2487  \\ 
    \hline
    IRDFT9-CZ & 36.9984 & 0.0887  \\
    \hline
\end{tabular} 
\end{center}
\end{table}

\begin{figure}[!tp]
\centering
\label{fig:differences}

\subfloat[MA50]{
\includegraphics[trim = 4cm 1cm 1cm 1cm, clip, width = 8cm]{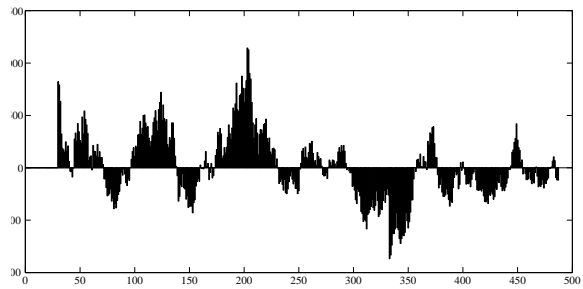} \label{MA_N}}

\subfloat[IRDFT9-LH]{
\includegraphics[trim = 4cm 1cm 1cm 1cm, clip, width = 8cm]{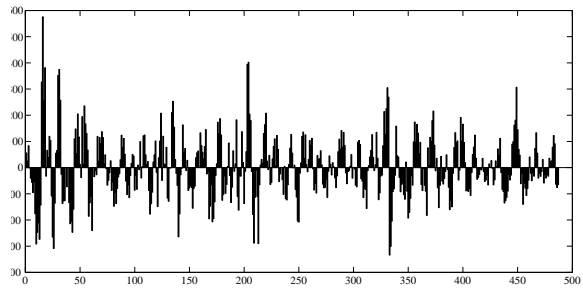} \label{LI_N}}

\subfloat[IRDFT9-CH]{
\includegraphics[trim = 4cm 1cm 1cm 1cm, clip, width = 8cm]{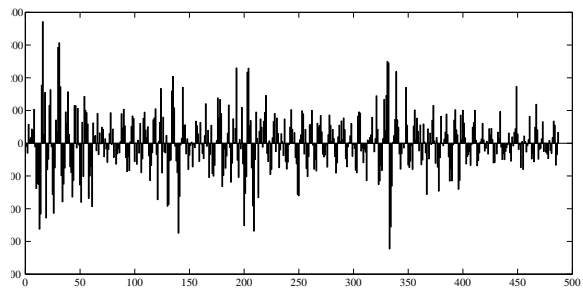} \label{CI_N}}

\subfloat[IRDFT9-LZ]{
\includegraphics[trim = 4cm 1cm 1cm 1cm, clip, width = 8cm]{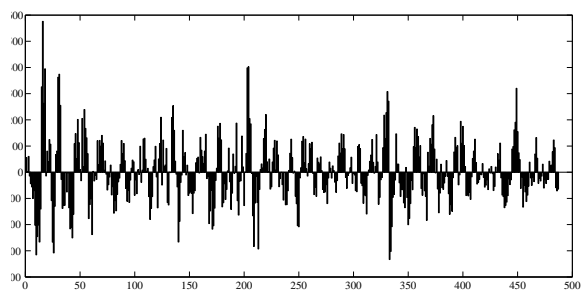} \label{CLI_N}}

\subfloat[IRDFT9-CZ]{
\includegraphics[trim = 4cm 1cm 1cm 1cm, clip, width = 8cm]{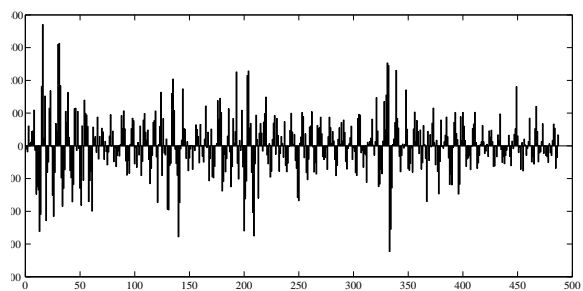} \label{CCI_N}}
\caption{Bar plot of point-wise differences between smoothed series and NIFTY50}
\end{figure}

In order to offer a quantitative comparison of the different series we considered the volatility of smoothed series (i.e., the standard deviation of daily returns), as a desired effect of smoothing is to reduce the variance of first order differences. The other metric is the mean absolute error (MAE) aimed at measuring the deviation between the smoothed and the original series due to the lag. Larger lags lead to larger errors. Results are reported in Table~\ref{tab:DR_Val}. They outline that F-transform smoothing provides a series that is closer to the original one. MA30 shows a larger value of MAE. As expected, if we look at MAE, a better fit is obtained by means of higher degree polynomials, i.e. cubic regression (IRDFT9-CH and IRDFT9-CZ) fits better than linear regression (IRDFT9-LH and IRDFT9-LZ). Obviously, this is payed in terms of higher standard deviation. No relevant differences are reported by considering hat-shaped (IRDFT9-LH and IRDFT9-CH) versus z-shaped basic functions (IRDFT9-LZ and IRDFT9-CZ). These considerations are visually summarized by Figure~\ref{fig:differences}, where we plot the point-wise differences between the smoothed series and the original NIFTY50 series.

\section{Conclusions}\label{sec:conc}
In this paper we proposed a generalization of F--transform that is driven by piecewise regression models, each associated to a specific set in the partition. We applied this approach to the smoothing of time series in finance, proving that the output series is better centered than moving average that is generally employed for this task.

\newpage




%

\bibliographystyle{myIEEETran}
\bibliography{references.bib}

\end{document}